\documentclass[conference]{IEEEtran}
\IEEEoverridecommandlockouts

\usepackage{cite}
\usepackage{amsmath,amssymb,amsfonts}
\usepackage{algorithmic}
\usepackage{graphicx}
\usepackage{textcomp}
\usepackage{xcolor}
\usepackage{listings}
\usepackage{booktabs}
\usepackage{balance}

\def\BibTeX{{\rm B\kern-.05em{\sc i\kern-.025em b}\kern-.08em
    T\kern-.1667em\lower.7ex\hbox{E}\kern-.125emX}}
\begin{document}

\title{An Improved Phase Coding Audio Steganography Algorithm}

\author{\IEEEauthorblockN{Guang Yang}
\IEEEauthorblockA{\textit{School of Information} \\
\textit{University of California, Berkeley}\\
Berkeley, CA, USA \\
guangyang19@berkeley.edu}
}

\maketitle

\begin{abstract}
Advances in speech synthesis have made voice cloning inexpensive and
convincing, and fraud built on synthetic audio is now a practical concern.
Embedding verifiable information directly in an audio signal is one response to
this problem. This paper revisits phase coding, a classical audio steganography
method that modifies the phase spectrum of a carrier. Traditional phase coding
places the entire payload in the first segment of the signal and propagates the
resulting phase difference through the remaining segments, which limits capacity
and degrades audio quality. We describe a segment-distributed variant that
spreads the payload across every segment and updates each segment
independently, and we pair it with a framing layer that adds a synchronization
word, a length field, a CRC-16 checksum, and Hamming(7,4) error correction.
We identify and correct a quantization defect that causes the method to fail on
speech, where a magnitude floor referenced to the whole signal is required for
the embedded phase to survive conversion to 16-bit samples. Measured over four
carriers, including real speech, the method recovers every message on a clean
channel while the classical baseline recovers none of the speech messages, and
it improves the stego signal-to-noise ratio by approximately 24~dB. Usable
capacity rises from 1023 bits to 32768 bits on a five second carrier at
equivalent embedding time. Under amplitude scaling and moderate requantization
the framing layer raises message recovery from 75\% to 100\%. We also report
the limits of the method. Because the payload occupies a narrow band below the
Nyquist frequency, it does not survive lossy compression, resampling, or
broadband additive noise, and the fixed bin placement offers no resistance to a
blind detector. We state these boundaries explicitly and outline keyed bin
selection as the path to addressing them.
\end{abstract}

\begin{IEEEkeywords}
Audio steganography, phase coding, Fast Fourier Transform, error correcting
codes, bit error rate, digital audio watermarking.
\end{IEEEkeywords}

\section{Introduction}

In recent years, cases of fraud involving AI-generated audio have become
increasingly common. In 2021, criminals used a cloned voice to impersonate a
company director and redirect a transfer of approximately \$35 million
\cite{brewster2021}. Incidents of this kind expose security weaknesses in
current audio communication, particularly in identity verification, digital
audio watermarking, and the validation of audio authenticity. Ensuring the
authenticity and integrity of audio content is therefore a pressing research
problem.

Two broad strategies address this problem. Passive detection trains a
classifier to distinguish synthetic speech from genuine recordings, an approach
represented by the ASVspoof benchmark series \cite{liu2023asvspoof} and
surveyed comprehensively by Pham et al. \cite{pham2025deepfake}. Proactive
provenance takes the opposite approach and embeds a verifiable mark in the
audio at the point of creation, so that authenticity can be checked later
without relying on a detector generalizing to unseen synthesis methods
\cite{sanroman2024audioseal, liu2024timbre}. This paper concerns the second
strategy.

Embedding information into audio through steganography is a direct route to
proactive provenance. Common audio steganography techniques include Least
Significant Bit (LSB) embedding, Echo Hiding, Phase Coding, and Spread Spectrum
\cite{kaur2014survey, singh2016comparative}. Among these, phase coding has the
advantage of a small effect on perceived audio quality, because the human
auditory system is relatively insensitive to absolute phase.

Phase coding hides information by modifying the phase components of an audio
signal \cite{bender1996}. The method applies the Fast Fourier Transform (FFT)
to convert the signal into the frequency domain, embeds the secret data in the
phase spectrum, and restores the signal to the time domain with the Inverse
Fast Fourier Transform (IFFT).

Traditional phase coding has three practical drawbacks. First, it computes and
propagates a phase difference across segments, which requires an additional
pass over the data. Second, the payload is concentrated in the first segment,
so the modification is localized and the capacity is bounded by one segment
rather than by the length of the carrier. Third, the raw payload bits are
transmitted without protection, so a single flipped bit corrupts the recovered
message and the receiver has no way to know that this has happened.

This paper addresses those three drawbacks. We describe a segment-distributed
variant of phase coding, add a framing layer that provides synchronization,
length signaling, integrity checking, and error correction, and evaluate both
against a classical baseline on four carriers. We also report where the method
fails, and we explain why those failures follow from the choice of embedding
band rather than from an implementation error.

\section{Related Work}

\subsection{Foundations}

The modern treatment of data hiding in audio begins with Bender et al.
\cite{bender1996}, which introduced phase coding alongside LSB coding and
spread spectrum methods, and with the companion work on echo hiding
\cite{gruhl1996echo}, which embeds data in the impulse response rather than the
spectrum. Cox et al. \cite{cox1997spread} established spread spectrum
watermarking as a robust alternative that trades capacity for resilience.
Petitcolas et al. \cite{petitcolas1999survey} provide the terminology that
separates steganography, watermarking, and covert channels, a distinction this
paper follows: our goal is a recoverable payload rather than an
adversarially robust mark.

\subsection{Time and transform domain embedding}

LSB embedding offers the highest raw capacity and the least robustness. Cvejic
and Seppanen raised LSB capacity by embedding in higher bit planes with
perceptual shaping \cite{cvejic2002capacity} and later moved the same idea into
the wavelet domain \cite{cvejic2002wavelet}. Transform-domain methods generally
trade capacity for resilience to filtering and resampling, a trade this paper
examines directly in Section~\ref{sec:evaluation}.

\subsection{Phase coding and its successors}

Phase coding has continued to develop since its introduction. Alsabhany et al.
\cite{alsabhany2019adaptive} proposed adaptive multi-level phase coding, which
selects an embedding level per operation and introduces interval centering
quantization to balance capacity, transparency, and robustness. More recently,
Yojith and Sethi \cite{yojith2026multibin} described multi-bin phase coding
with adaptive post-processing for audio-in-audio hiding. Both distribute the
payload over more of the spectrum than the original single-segment method, as
does the approach described here. The distinguishing feature of our work is the
explicit framing layer, which separates the channel coding concern from the
embedding geometry.

\subsection{Steganalysis}

Any embedding scheme must be assessed against detection. Zeng et al.
\cite{zeng2007steganalysis} give a steganalysis algorithm targeted specifically
at phase coding, exploiting the concentrated phase modification that
single-segment embedding produces. Ghasemzadeh and Kayvanrad
\cite{ghasemzadeh2018review} survey the field more broadly. Detection has since
moved to learned models: Chen et al. \cite{chen2017cnn} showed that a
convolutional network outperforms hand-crafted features on audio steganalysis.
Section~\ref{sec:limitations} reports a direct measurement of how the present
implementation fares against a simple blind detector, and the result is not
favorable.

\subsection{Error correction in information hiding}

Coding theory has long been applied to data hiding. Fridrich and Soukal
\cite{fridrich2006matrix} developed matrix embedding, which uses linear codes,
Hamming codes among them, to minimize the number of carrier changes required
for a given payload. Their objective is embedding efficiency. The framing layer
in this paper uses the same code family for a different purpose, namely
correcting errors introduced by the channel after embedding. Raiyan and Kabir
\cite{raiyan2025screedsolo} take an approach structurally similar to ours in
the image domain, combining an LSB payload with Reed--Solomon coding for
robustness.

\subsection{Learned audio watermarking}

Recent watermarking systems learn the embedding and extraction functions
jointly rather than designing them by hand. DeAR \cite{liu2023dear} trains
against a simulated re-recording channel and survives playback and capture
over the air. IDEAW \cite{li2024ideaw} uses an invertible network with a dual
embedding stage that separates message payload from locating information, a
division of labor comparable in spirit to the framing layer described in
Section~V, though realized by learned components rather than an explicit code.
Cui et al. \cite{cui2020multistage} extend the same idea to a larger payload,
hiding images in audio through multi-stage residual embedding.

These systems obtain robustness by learning a channel model, which is
effective against distortions that can be simulated during training. The method
described here obtains a narrower form of robustness from an explicit and
inspectable code, and it runs in milliseconds on a CPU with no trained model
and no accelerator. The two approaches suit different deployment constraints
rather than competing directly.

\section{Algorithm}

The improved algorithm proceeds in five steps and distributes the embedded
information across the entire audio signal.

\subsection{Segmentation}

The input signal $S$ is divided into $n$ contiguous segments $S_i$, with
$i \in \{1, \dots, n\}$, each of length $l$. The segment length is derived from
the payload size $m$ in bits as
\[
l = 2 \cdot 2^{\lceil \log_2 (2m) \rceil},
\]
and $n = \lfloor |S| / l \rfloor$. Samples beyond $n \cdot l$ are left
unmodified.

\subsection{Spectral decomposition}

For each segment the FFT yields a magnitude and a phase,
\[
A_i = |\mathrm{FFT}(S_i)|, \qquad \phi_i = \angle \mathrm{FFT}(S_i).
\]

\subsection{Phase assignment}

Each payload bit $d_j$ maps to a phase value in quadrature,
\[
\phi_{\mathrm{data}}[j] =
\begin{cases}
+\pi/2, & d_j = 0, \\
-\pi/2, & d_j = 1,
\end{cases}
\]
so that the detector recovers a bit by testing the sign of the phase. Writing
the two states as far from the decision boundary as the phase range allows
maximizes the margin against perturbation.

Let $c = l/2$ denote the Nyquist bin of a segment and let $g$ be a guard of
bins left untouched at the band edge. Segment $i$ carries the bits indexed
$[\,i m / n,\ (i+1) m / n\,)$, written to the $k$ bins below the guard,
\[
\phi'_i[c - g - k + j] = \phi_{\mathrm{data}}[j], \quad j = 1, \dots, k .
\]
Conjugate symmetry is enforced so the inverse transform is real,
\[
\phi'_i[c + g + j] = -\phi_{\mathrm{data}}[k - j + 1].
\]
The residual imaginary component after reconstruction is on the order of
$10^{-12}$, which is floating point noise.

\subsection{Magnitude floor}
\label{sec:floor}

Phase carries information only insofar as the corresponding magnitude survives
the return to integer samples. Reconstructed samples are rounded to 16-bit
integers, and rounding perturbs the phase of a bin in proportion to the ratio
between the quantization step and the bin magnitude. Where a payload bin is
weak, that perturbation is large enough to flip the sign that the detector
reads.

This is not a hypothetical concern. Speech spectra roll off steeply, so the
bins just below Nyquist carry very little energy, and speech also contains
extended near-silent passages. In our measurements the payload bins of the
speech carrier had a mean magnitude of 1629 against 5154 to 14824 for the other
carriers, and 28.4\% of speech segments fell below 5\% of the peak segment
RMS. The bit error rate on speech was 0.095 to 0.119 while the synthetic
carriers were nearly error free.

The remedy is a lower bound on the magnitude of payload bins,
\[
A'_i[b] = \max \left( A_i[b],\ \alpha \cdot \overline{A} \right),
\]
where $b$ ranges over the payload bins of segment $i$ and $\overline{A}$ is the
mean magnitude over all segments of the signal.

The reference quantity matters. A floor computed from each segment's own mean
scales down with a quiet segment and therefore provides no protection precisely
where protection is needed. Measured on the speech carrier, a per-segment floor
reduced the bit error rate only from 0.119 to 0.091 and cost 3.8~dB of stego
SNR. A floor referenced to the whole signal, with $\alpha = 0.02$, reduces the
bit error rate to zero on all four carriers at a cost below 0.05~dB.

\subsection{Reconstruction}

Each segment is reconstructed as
\[
S'_i = \mathrm{IFFT}\left( A'_i \cdot \exp(j \phi'_i) \right),
\]
and the segments are concatenated. Samples are clipped to the representable
range rather than cast directly, since a direct cast wraps on overflow and
turns a sample slightly above full scale into a large negative value, producing
an audible click.

Unlike the traditional method, no phase difference is propagated between
segments. Each segment is updated independently, which removes the second pass
over the data and prevents distortion introduced in one segment from
propagating through the rest of the signal.

\section{Framing Layer}

The embedding described above transmits raw bits. A single flipped bit
corrupts the recovered message, the receiver cannot tell that the message is
damaged, and the payload length must be communicated separately because the
segment geometry cannot be derived without it.

A framing layer closes these gaps. Before channel coding, a frame is

\[
\underbrace{\texttt{SYNC}}_{16} \;\|\;
\underbrace{\texttt{LEN}}_{16} \;\|\;
\underbrace{\texttt{PAYLOAD}}_{8\nu} \;\|\;
\underbrace{\texttt{CRC}}_{16}
\]

where \texttt{SYNC} is a fixed synchronization word, \texttt{LEN} is the
payload length in bytes, and \texttt{CRC} is a CRC-16/CCITT-FALSE checksum over
the payload. Hamming(7,4) is applied to the concatenation, expanding the frame
by a factor of $7/4$ and correcting one bit error in each seven-bit codeword.
An optional repetition code with odd factor $r$ and majority voting may be
applied afterward.

The three mechanisms serve distinct purposes and are worth separating clearly.
Hamming(7,4) repairs isolated errors. CRC-16 detects corruption that Hamming
cannot repair, converting a silent failure into a reported one. Repetition
trades capacity for margin. None of them can recover a payload from a channel
that carries no information, a point Section~\ref{sec:limitations} returns to.

\section{Evaluation}
\label{sec:evaluation}

\subsection{Experimental setup}

Four carriers were used, each five seconds long, mono, 44.1~kHz, 16-bit: a pure
440~Hz tone, a three-note chord, band-limited noise, and a real speech
recording. The four span sparse and dense spectra, periodic and aperiodic
signals, and the intended application.

Four schemes are compared. \emph{Classical} is Bender's single-segment method
with phase difference propagation, serving as the baseline. \emph{Improved} is
the segment-distributed method of Section III without channel coding.
\emph{Improved + H(7,4)} adds the framing layer with Hamming coding.
\emph{Improved + H(7,4) + R3} adds threefold repetition.

Each configuration was run over 20 independent trials with random payloads.
Reported values are means, and error bars denote one standard deviation. All
measurements were produced on an AMD Ryzen 9 9950X.

\subsection{Clean channel recovery}

\begin{figure}[tb]
    \centering
    \includegraphics[width=\linewidth]{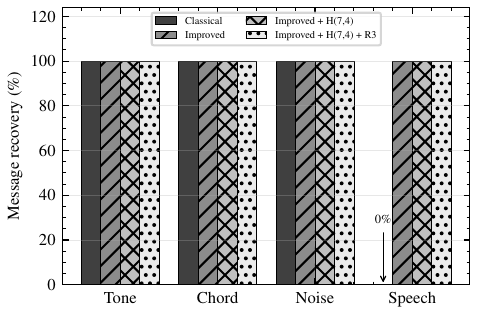}
    \caption{Message recovery on a clean channel. The classical baseline
    recovers no messages at all from the speech carrier, while every
    segment-distributed variant recovers all of them.}
    \label{fig:clean}
\end{figure}

Figure~\ref{fig:clean} reports the fraction of messages recovered exactly with
no channel distortion.

The segment-distributed method recovers every message on every carrier. The
classical baseline recovers every message on the three synthetic carriers and
none on speech. Averaged over all carriers and payload sizes, the baseline
reaches 75.0\% message recovery with a raw bit error rate of 0.0457, against
100\% and 0.0000 for the improved method.

The speech result is the more informative of the two. A method that fails
completely on the signal class it is intended for is not usable, and the
failure is invisible to an evaluation that uses only synthetic carriers. The
magnitude floor of Section~\ref{sec:floor} is what closes this gap.

\subsection{Perceptual transparency}

\begin{figure}[tb]
    \centering
    \includegraphics[width=\linewidth]{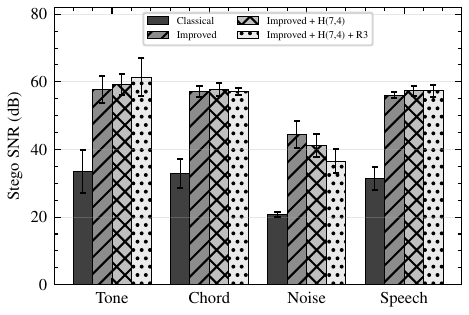}
    \caption{Stego signal-to-noise ratio by carrier. Higher is better. The
    segment-distributed method improves transparency by roughly 24~dB relative
    to the classical baseline.}
    \label{fig:snr}
\end{figure}

Figure~\ref{fig:snr} reports the signal-to-noise ratio between the carrier and
the stego signal.

The improvement is substantial and consistent. On speech the SNR rises from
31.4~dB to 56.1~dB, on the tone from 33.4~dB to 57.7~dB, and on the chord from
32.9~dB to 57.2~dB. The noise carrier shows the smallest margin, 20.8~dB
against 44.5~dB, which is expected because a dense spectrum offers less room
for a modification to hide.

The mechanism is straightforward. The classical method concentrates the entire
payload in one segment and then rotates every remaining segment by the
resulting phase difference, so a large modification in one place is spread over
the whole signal. Distributing the payload means each segment carries a few
bits and no propagation step is required.

\begin{figure}[tb]
    \centering
    \includegraphics[width=\linewidth]{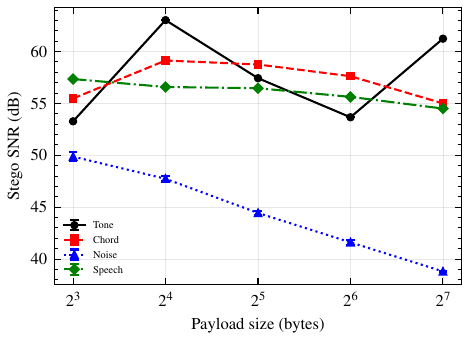}
    \caption{Stego SNR against payload size for the segment-distributed method.
    Transparency remains above 38~dB across the tested range.}
    \label{fig:payload}
\end{figure}

Figure~\ref{fig:payload} shows how transparency varies with payload size.

\subsection{Robustness to benign distortion}

\begin{figure}[tb]
    \centering
    \includegraphics[width=\linewidth]{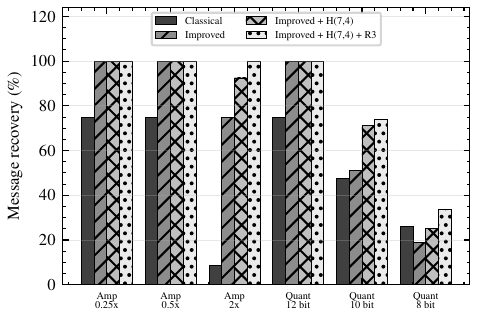}
    \caption{Message recovery under amplitude scaling and requantization. The
    amplitude 2x group shows the clearest coding gain, rising from 75\% to
    100\% as channel coding is added.}
    \label{fig:benign}
\end{figure}

Figure~\ref{fig:benign} reports message recovery under distortions that
preserve the embedding band.

Amplitude scaling by 0.25 and 0.5 is recovered perfectly by every
segment-distributed variant, against 75.0\% for the baseline. Amplitude scaling
by 2.0 introduces clipping, and here the value of the framing layer is
explicit: recovery rises from 75.0\% uncoded to 92.5\% with Hamming coding and
100\% with repetition added. Requantization to 12 bits is recovered perfectly
by all improved variants.

Two results run the other way and are reported for completeness. At 10-bit
requantization recovery falls to 51.2\% uncoded, improving to 71.2\% and 73.8\%
with coding. At 8 bits the uncoded improved method reaches 18.8\%, below the
classical baseline at 26.2\%, and coding recovers only part of the difference.
Aggressive requantization raises the noise floor across the whole spectrum,
including the narrow band the payload occupies, and the magnitude floor of
Section~\ref{sec:floor} is calibrated for 16-bit output rather than for a
severely reduced bit depth.

\subsection{Capacity and computational cost}

\begin{figure}[tb]
    \centering
    \includegraphics[width=\linewidth]{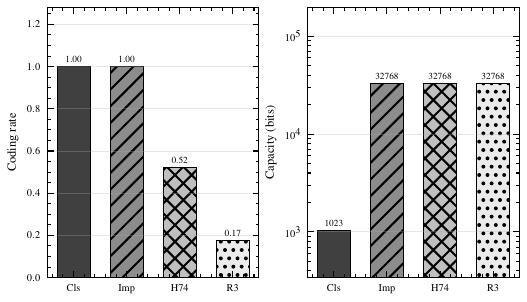}
    \caption{Effective coding rate and usable capacity on a five second
    carrier. Cls is classical, Imp is improved, H74 adds Hamming(7,4), and R3
    adds threefold repetition.}
    \label{fig:capacity}
\end{figure}

Figure~\ref{fig:capacity} reports the effective coding rate and the usable
capacity of each scheme.

Usable capacity rises from 1023 bits for the classical method to 32768 bits for
the segment-distributed method, a factor of 32, because capacity is bounded by
the whole carrier rather than by a single segment. Embedding a 64-byte payload
takes 5.1~ms against 5.5~ms for the baseline, so the capacity gain does not
cost time. Extraction takes 1.2~ms.

Channel coding reduces the effective rate from 1.000 to 0.522 with Hamming
coding and to 0.174 with threefold repetition added, while leaving runtime
essentially unchanged at 5.0 and 5.4~ms. The rate cost is the price of the
recovery gains reported above, and the choice between the variants is an
application decision rather than a technical one.

\section{Limitations}
\label{sec:limitations}

The measurements above establish what the method does well. This section
reports where it fails. Each result is measured, and each follows from the same
design decision: the payload occupies a narrow band immediately below the
Nyquist frequency.

\subsection{Additive noise}

\begin{figure}[tb]
    \centering
    \includegraphics[width=\linewidth]{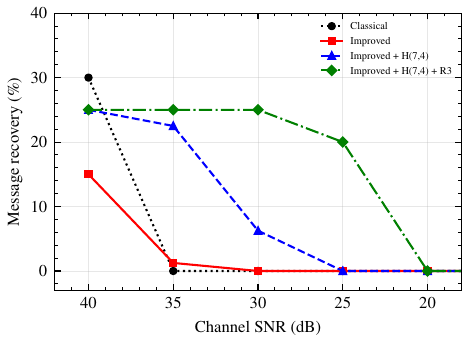}
    \caption{Message recovery under additive white Gaussian noise. Coding gain
    is visible in the ordering of the variants, but no configuration is usable
    below 25~dB channel SNR.}
    \label{fig:awgn}
\end{figure}

Figure~\ref{fig:awgn} reports message recovery under additive white Gaussian
noise.

The method does not resist broadband noise. At 40~dB channel SNR recovery is
between 15\% and 30\%, and at 20~dB and below every configuration recovers
nothing.

The cause is a mismatch of bandwidths rather than a coding weakness. Noise
scaled to the power of the whole signal distributes energy across all bins,
while the payload occupies only a few. Measuring the local signal-to-noise
ratio within the payload band makes this concrete: at a nominal channel SNR of
40~dB the payload band SNR is 5.0~dB for the tone carrier and 2.4~dB for
speech, and at a nominal 30~dB it is already negative, at $-5.0$~dB and
$-7.6$~dB respectively. A bin whose magnitude falls below the per-bin noise
level carries no recoverable phase, and no channel code repairs a channel that
carries no information.

What the figure does show is coding gain within the region where recovery is
non-zero. At 30~dB the uncoded method recovers nothing while Hamming coding
recovers 6.2\% and repetition recovers 25.0\%. The ordering is consistent and
confirms that the framing layer functions as intended, but it does not support
a general claim of noise robustness.

\subsection{Lossy compression and resampling}

The payload does not survive lossy compression or sample rate conversion. MP3
encoding at 128~kbps yields a bit error rate of 0.504, at 64~kbps 0.461, and
resampling from 44.1~kHz to 22.05~kHz and back yields 0.500. These values are
at the level of random guessing, which is to say the payload is destroyed
rather than degraded.

This follows directly from the embedding band. Perceptual codecs discard
content near the Nyquist frequency because it contributes least to perceived
quality, and downsampling to 22.05~kHz removes that band entirely. Moving the
payload to a genuine mid-frequency band, specified in Hz rather than as an
offset from the Nyquist bin, is the necessary remedy and is a design change
rather than a parameter adjustment.

\subsection{Detectability}

The embedding position in the present implementation is deterministic. Payload
bins occupy $k$ consecutive positions below a fixed guard, and both the guard
and the segment length are derived from public quantities. There is no secret
parameter.

We measured the consequence with a blind detector given only the stego signal
and the general knowledge that phase coding writes values near $\pm \pi/2$. The
detection statistic is the fraction of segments whose phase at a candidate bin
falls within 0.05 radians of quadrature. Scanning five candidate segment
lengths against five candidate bin offsets, 25 combinations in total, the
statistic at the true bin rose from 0.009 on the clean carrier to 0.791 on the
stego signal, a factor of 84, and identified the correct segment length and bin
exactly. Using only the recovered parameters, the payload was then extracted
with a bit agreement of 1.000.

We therefore make no claim of improved undetectability. Distributing the
payload across segments is a genuine advantage in capacity and perceptual
quality, as Section~\ref{sec:evaluation} shows, but distribution across
segments at a fixed spectral position is not concealment.

The remedy is keyed pseudorandom bin selection. If payload positions were drawn
from a generator seeded with a shared secret, an attacker without the key would
face a search over the key space rather than a 25 point scan, and the detection
statistic would be diluted across many candidate bins instead of concentrated
in one. This is the principal direction for future work.

\subsection{Framing layer integration}

The framing layer is implemented and verified, but it is not yet wired into the
public embedding interface, which still requires the payload length as an
argument. Recovering the length from the frame requires a segment geometry that
does not itself depend on the payload size, which in turn requires decoupling
the segment length from the payload. That change is straightforward and is
planned.

\section{Conclusion}

This paper revisited phase coding audio steganography and addressed three
practical weaknesses of the traditional method. Distributing the payload across
every segment and updating each segment independently removes the phase
difference propagation pass, raises usable capacity from 1023 bits to 32768
bits on a five second carrier, and improves the stego signal-to-noise ratio by
approximately 24~dB. A framing layer supplying synchronization, a length field,
CRC-16 integrity checking, and Hamming(7,4) correction converts silent
corruption into detected corruption and raises message recovery under clipping
from 75\% to 100\%.

We also identified a quantization defect that made the method unusable on
speech, the signal class of greatest practical interest, and traced it to
payload bins whose magnitude is too small for the embedded phase to survive
conversion to 16-bit samples. A magnitude floor referenced to the whole signal,
rather than to each segment, reduces the bit error rate on speech from 0.119 to
zero at a cost below 0.05~dB of transparency.

The method has clear boundaries, and we have stated them with the same
measurements that establish its strengths. The payload does not survive lossy
compression, resampling, or broadband additive noise, and its fixed spectral
position offers no resistance to a blind detector that scans for phase
anomalies. All three limitations trace to the choice of embedding band and the
absence of a secret parameter. Future work will move the payload to a
mid-frequency band specified in absolute terms and introduce keyed
pseudorandom bin selection, which together address robustness and detectability
at their common source.

\balance

\end{document}